# Reduced radiative conductivity of high and low spin FeO$_6$-octahedra in the Earth's lower mantle

Sergey S. Lobanov[1,2,3,*], Nicholas Holtgrewe[1,4], Alexander F. Goncharov[1,2,5]

[1]Geophysical Laboratory, Carnegie Institution of Washington, Washington, DC 20015, USA

[2]Key Laboratory of Materials Physics, Institute of Solid State Physics CAS, Hefei 230031, China

[3]V.S. Sobolev Institute of Geology and Mineralogy SB RAS, Novosibirsk 630090, Russia

[4]Howard University, 2400 Sixth Street NW, Washington, DC 20059, USA

[5]University of Science and Technology of China, Hefei 230026, China

*slobanov@carnegiescience.edu

## Abstract

The ability of Earth's mantle to conduct heat by radiation is determined by optical properties of mantle phases. Optical properties of mantle minerals at high pressure are accessible through diamond anvil cell experiments, but because of the extensive thermal radiation at T > 1000 K such studies are limited to lower temperatures making it necessary to model the radiative thermal conductivity at mantle conditions. Particularly uncertain is the temperature-dependence of optical properties of lower mantle minerals across the spin transition as the spin state itself is a strong function of temperature. Here we use laser-heated diamond anvil cells combined with a pulsed ultra-bright supercontinuum laser probe and a synchronized time-gated detector to examine optical properties of high and low spin ferrous iron at 45-73 GPa and to 1600 K in an octahedral crystallographic unit (FeO$_6$), one of the most abundant building blocks in the mantle. Siderite (FeCO$_3$) is used as a model for FeO$_6$-octahedra as it contains no ferric iron and exhibits a sharp optically apparent pressure-induced spin transition at 44 GPa, simplifying data interpretation. We find that the optical absorbance of low spin FeO$_6$ is substantially increased at 1000-1200 K due to the partially lifted Laporte selection rule. The temperature-induced low-to-high spin transition, however, results in a dramatic drop in absorbance of the $FeO_6$ unit in siderite. The absorption edge (Fe-O charge transfer) red-shifts (~ 1 cm$^{-1}$/K) with increasing temperature and at T > 1600 K becomes the dominant absorption mechanism in the visible range, suggesting its superior role in reducing the ability of mantle minerals to conduct heat by radiation. This implies that the radiative thermal conductivity of analogous $FeO_6$-bearing minerals such as ferropericlase, the second most abundant mineral in the Earth's lower mantle, is substantially reduced approaching the core-mantle boundary conditions. Finally, our results emphasize that optical properties of mantle minerals probed at room temperature are insufficient to model radiative thermal conductivity of planetary interiors.

**Keywords:** Optical properties; thermal conductivity; spin transition; high pressure; siderite; ferropericlase

## Introduction

Heat conduction through the Earth largely determines planetary internal structure[1-3]. Therefore, knowledge of mineral thermal conductivity under relevant mantle pressures and temperatures is vital for our understanding of the Earth's interior. Radiative thermal conductivity ($k_{rad}$) is believed to play an increasingly important role at T > 800 K (Ref.[4-6]) because of the near-cubic dependence of $k_{rad}$ on temperature[3]. Absolute values of $k_{rad}$ are determined by absorbance in the visible and near infrared (IR) range[7-9] provided that the mineral grain size in the mantle is sufficiently large[10]; this is why the initial proposal of the increasing importance of heat conduction by radiation with depth[4] had triggered interest in optical properties of Fe-bearing olivine and pyroxene[7], dominant upper mantle phases. Soon it was realized that optical properties of minerals themselves are highly dependent on temperature and pressure; thus, these must be addressed at relevant mantle conditions. Diamond anvil cells (DACs) have previously been employed to study variations in optical and near IR absorption at high pressure, but at room temperature[11-19]. Likewise, high temperature studies at near-ambient pressures were not conclusive[8, 9, 20-22]. More recently, examining optical properties of minerals in resistively-heated DACs helped understand the combined effect of pressures and temperature, but the results are scarce and limited to T < 900 K (Ref.[11, 23]).

In the absence of spectroscopic data at simultaneous conditions of high pressure and temperature a common approach has been to use room temperature absorption coefficients to evaluate $k_{rad}$ values as a function of temperature[13, 15, 16, 18, 24]. Experimental measurements of optical properties of the major phases at lower mantle P-T conditions (P > 24 GPa, T > 2000 K) are needed to test the validity of this traditional approach and to gain accurate values of thermal conductivity.

Absorption spectra of bridgmanite and ferropericlase (FP), the major lower mantle phases, are governed by the crystal field (CF), iron-oxygen, and $Fe^{2+}$-$Fe^{3+}$ transitions[25]. Unfortunately, these bands often overlap making it difficult to isolate the contribution of a specific crystallographic environment to the optical properties. In addition, optical properties are sensitive to the spin state of iron-bearing minerals and a reduced radiative conductivity scenario has been proposed for the low spin (LS) mantle[11, 14, 16]. On top of this, high temperature is expected to affect mineral optical properties[25], but the effect of it has been unclear since optical absorption experiments at T > 1000 K are very challenging as the probe signal vanishes in the vast blackbody radiation. The course of this work is focused on overcoming the experimental barriers by combining laser-heated (LH) DACs with an ultra-bright probe synchronized with a time-gated detector.

Here we investigate optical properties of the $FeO_6$-octahedron, one of the most common building blocks of mantle minerals, at P = 45-85 GPa and T = 300-1625 K in the LS state and across the low-to-high spin transition. We chose siderite ($FeCO_3$) as a model to access optical properties of the FeO$_6$-octahedron in FP with only ferrous iron because siderite contains no ferric iron. Conveniently for the direct comparison, both minerals undergo a spin transition at very similar pressures: 44 GPa (siderite)[26-30] and at P > 45-65 GPa (FP)[31-34]. Our results show for the first time that minerals with LS ferrous iron have a substantially increased intrinsic absorption in the visible and near IR spectral range at high temperatures, suggesting a greatly reduced radiative conductivity. This result underscores the necessity of measurements of optical properties at simultaneous conditions of high pressure and temperature to determine $k_{rad}$ of lower mantle minerals.

**Experimental Methods**

**Samples.** Natural siderite (Panasqueira tungsten mine, Covilhã, Castelo Branco, Portugal) used in this study is identical to that used in our previous report on siderite optical properties at high pressure and room temperature (Ref.[26]). Chemical composition (Fe$_{0.95}$,Mn$_{0.05}$CO$_3$), homogeneity, and low impurity content (< 1 at.%) of our samples were confirmed in the Geophysical Laboratory using a JEOL JSM-6500F field emission scanning electron microscope. Flat, rhombohedrally cleaved single crystals (typically 30 x 50 x 10 µm$^3$) without observable defects were selected under an optical microscope. Subsequently, the crystals were positioned in the DAC sample cavity (70-120 µm in diameter) between two dried KCl or CaF$_2$ wafers (10-15 µm thick) serving as a thermal insulation and pressure medium. The ruby R-1 fluorescence method[35], the position of diamond (anvil) Raman edge[36] and siderite CF absorption band[26] were used to evaluate pressure. Pressure uncertainty was on the order of 3-4 GPa and reflects typical differences in apparent pressure using the 3 pressure gauges.

**Optical absorption in laser-heated diamond anvil cell.** The 1070 nm double-sided LH system has been described in detail[37]. Additionally, Leukos Pegasus pulsed supercontinuum (broadband) laser (400-2400 nm) serving as a light source for absorption measurements was introduced into the DAC using a broadband beamsplitter. The transmission of the supercontinuum was collected by a spectrometer with a synchronized iCCD detector (Andor iStar SR-303i-A), wavelength range 350-850 nm. The heating laser was modulated for 1 second, initiating the beginning of one spectral collection. At a specified delay (200 ms, sufficient to reach a steady temperature regime[38]), a train of supercontinuum pulses (4 ns pulse length, 250 kHz repetition rate) was collected at the gated iCCD detector (gate width 30 ns) operating at ~41 kHz for 500 ms (Fig. 1). A spectrum at each grating position (300 gr/mm) was a sum of all pulses accumulated on the iCCD over 500 ms. The cycle was repeated for each grating position and these spectra were stitched together for the final spectrum in the 12500-22500 cm$^{-1}$ (444-800 nm) spectral range. The timing of the system was important to overcome a poor signal-to-noise ratio due to background thermal emission from the heated sample. The sample absorbance was calculated as $A(v) = -\log_{10}(I_{sample}^{T} - Bckg^{T}/I_{reference}^{300K} - Bckg^{300K})$, where $I_{sample}^{T}$ is the intensity of light transmitted through the sample at temperature $T$, $I_{reference}^{300K}$ is the intensity of the radiation passed through the pressure medium at room temperature, $Bckg^{T}$ and $Bckg^{300K}$ are backgrounds at $T$ and at room temperature, respectively. Reflection from the sample (< 4 % of $I_{reference}^{300K}$) was neglected in absorption measurements as it is comparable to other experimental uncertainties[12].

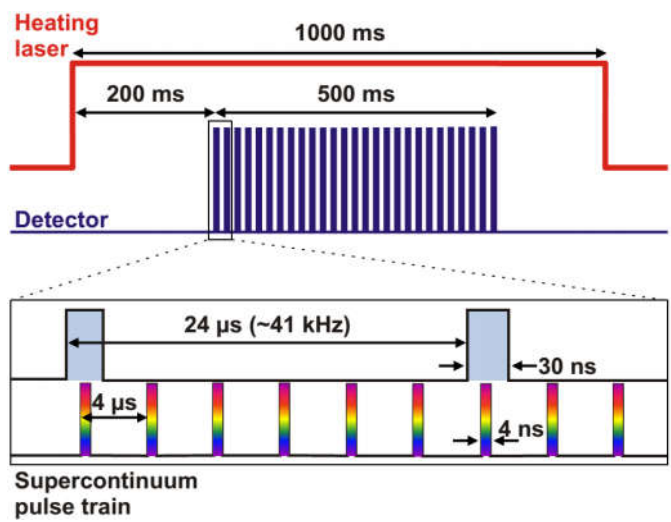

**Fig. 1.** Timing of the optical absorption measurements at high temperature. Every sixth pulse of the supercontinuum pulse train reaches the detector operated at 41 kHz.

For double-sided temperature measurements thermal radiation was imaged onto the iCCD detector using the full grid of the CCD array (grating centered at 700 nm), with emissions from both sides of the DAC simultaneously imaged and offset from one another. Temperature was measured spectroradiometrically with collection times ranging from 0.1 to 10 seconds depending on the emission intensity. The optical response of the system was calibrated using a standardized lamp (Optronics Laboratories OL 220C) and the emission spectrum was fitted to the Planck black body radiation function using the T-Rax software (C. Prescher) to extract the temperature value.

Typical collection runs involved: (i) optical absorption measurement before LH, (ii) temperature measurement, (iii) optical absorption measurement at high temperature, (iv) optical absorption measurement in quenched sample, and (v) temperature measurement. The second temperature measurement was to verify temperature consistency over the cycle. Typical differences between 2 measurements were < 50 K. Temperatures reported here are averaged over the 2 temperature collections and from both DAC sides. We assume the overall temperature uncertainty in our experiments to be typical of LH DAC temperature measurements by spectraradiometry (± 100-150 K)[39].

## Results and Discussion

At **P < 44 GPa** absorbance of HS siderite was identical to that reported for siderite in Ne pressure medium[26]. Briefly, HS siderite remains visually colorless and overall optical absorbance is small with only a weak CF band centered at ~ 14100 cm$^{-1}$ (43 GPa). This is why we could not couple the 1070 nm heating laser to siderite at P < 44 GPa and did not examine the HS phase at high temperature. Upon the spin transition in the relatively rigid KCl or CaF$_2$ pressure medium siderite turned visually black, having 0.3-0.5 higher absorbance as compared to the samples of similar thickness, but in the Ne pressure medium (Fig 2)[26], although the spectra are qualitatively similar.

**45 GPa**

The strong absorption band centered at ~15500 cm$^{-1}$ seen in the optical absorption spectra before and after laser-heating (Fig. 2) is characteristic of LS siderite and is attributed to the $^1A_{1g} \rightarrow {}^1T_{1g}$ transition in Fe$^{2+}$ electronic structure split by an octahedral CF[26]. At 1200 and 1300 K this absorption peak is no longer present in the spectrum, but substituted by a weak absorption band centered at ~14500 cm$^{-1}$, attributed to the $^5T_{2g} \rightarrow {}^5E_g$ transition in HS siderite. The band assignment was based on the energy and number of spin-allowed excitations for a HS and LS *d6* ion in an octahedral CF[26]. The abrupt change in the siderite optical absorption and correspondence with room temperature spectra of HS and LS siderite (black curves in Fig. 2) indicate that LS to HS transformation at 1200-1300 K has completed. The absorption edge (AE) red-shifts with increasing temperature which was previously observed in Fe-bearing silicates and attributed to the bang gap closure due to the electronic band broadening[20, 22, 25, 40]. Please note that the absorption spectra before and after LH are almost identical indicating a reversible character of the observed transformations in the heating cycle.

Heating to T > 1400-1500 K produced an opaque region in the heated spot which shows a complex x-ray diffraction pattern inconsistent with siderite. An isochemical phase transition in FeCO$_3$ at P > 40 GPa and T > 1400 K was reported by Liu et al.[41]. Alternatively, a decomposition to a complex mixture consisting of a tetrahedrally-coordinated carbonate phase (Fe$_4$(CO$_4$)$_3$), diamond, carbon monoxide, and high pressure form of Fe$_3$O$_4$ may be expected from the LH DAC experiments in FeO-CO$_2$ and FeCO$_3$ systems at P > 50 GPa (Refs.[42, 43]). Preliminary analysis of the x-ray diffraction measured in this work at similar conditions (P = 45-60 GPa) after LH to T > 1400-1500 K is consistent with the decomposition. Likewise, secondary electron microscopy analyses of recovered samples prepared by focused ion beam reveal a chemical inhomogeneity across the LH spot (Fig. 3) in favor of the decomposition scenario.

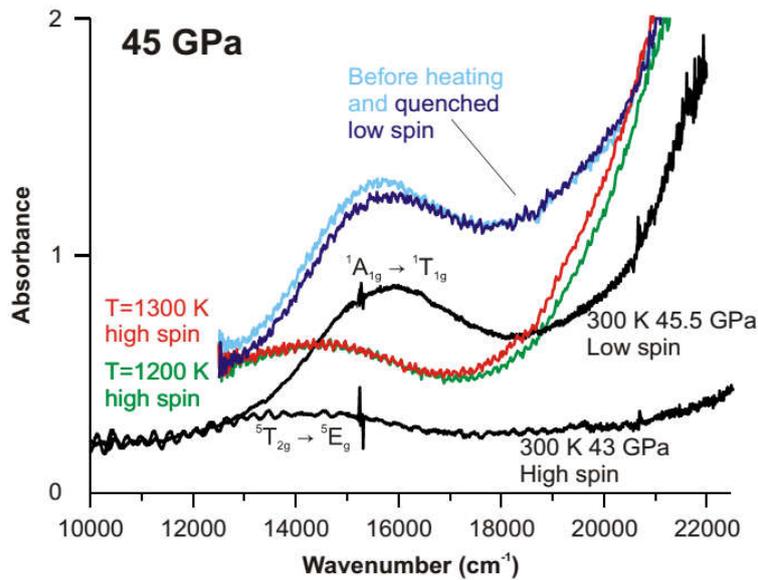

**Fig. 2.** High temperature siderite absorption spectra at 45 GPa. The spectra measured before heating and quenched after 1300 K are shown in light and dark blue, respectively. Green and red curves are absorption spectra at 1200 K and 1300 K, respectively. Spectra shown in black represent room temperature absorption data on HS (43 GPa) and LS (45.5 GPa) siderite after Lobanov et al. [26], shown for comparison. Note the temperature-induced red-shift of the AE (absorption edge). The difference in the absorbance magnitude is due to the sample damage introduced by the rigid pressure medium used in this study ($CaF_2$ vs Ne in Ref.[26])

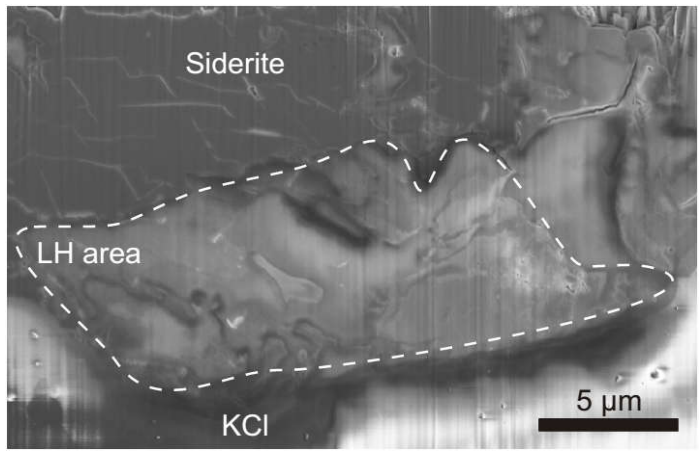

**Fig. 3.** Scanning electron microscope image of a siderite sample recovered after LH to T ~ 2000 K at P = 48 GPa. The sample was cut open with a focused ion beam. A homogeneous area in the upper left corner is unreacted siderite as confirmed by energy dispersive x-ray spectroscopy analysis. The area outlined by the white dashed curve roughly corresponds to the LH region and shows multiple microscale heterogeneities consistent with the siderite decomposition scenario. Bottom part of the micrograph almost exquisitely corresponds to KCl which served as a pressure medium and thermal insulation.

**55 GPa**

Upon heating to T ~ 1000 K at 55 GPa the $^1A_{1g} \rightarrow {}^1T_{1g}$ CF band broadens, red-shifts, and intensifies (Fig. 4A). Thermal broadening of CF bands results from the participation of excited vibrational states in the electron transfer[25]. The red-shift is likely due to the thermal expansion as the energy separation between the $t_{2g}$ and $e_g$ orbitals is inversely proportional to the fifth power of the metal-ligand distance[25]. Intensity variations of CF bands with temperature reported in the literature show a less regular behavior[8, 9, 11, 22, 23, 44, 45], but often may be understood in terms of temperature-induced changes in the CF symmetry[40] lifting the Laporte selection rule which impedes electron transfer among the same parity states (*e.g. d-d* transitions). Several mechanisms may weaken the Laporte selection rule. For example, an interaction of *d*-orbitals with thermally activated odd-parity vibrational states (termed vibronic coupling), has been proposed to explain the temperature-enhanced absorption of $Fe^{2+}$ CF bands in olivine[8, 21, 25]. Alternatively, distorting the centrosymmetric octahedral CF may lead to an increased mixing of iron *3d* and *4p* orbitals partly lifting the Laporte selection rule[25].

At 1200 K the intensity of the CF band has dropped, signaling an increased contribution of the HS states ($^5T_{2g} \rightarrow {}^5E_g$ transitions) which are less absorbing (Fig. 4B), consistent with the siderite spin state diagram[41]. At 1625 K it was no longer possible to reliably trace the CF band due to its progressive broadening and overlap with the red-shifted AE (Fig. 4C). All high temperature spectra show an increased absorption at the lower wavenumbers (12500-15000 cm$^{-1}$) compared to corresponding room temperature spectra, which might be due to the temperature-induced asymmetrical broadening and red-shift of the $^1A_{1g} \rightarrow {}^1T_{1g}$ band. Likewise, AE may contribute to the absorbance at 12500-15000 cm$^{-1}$ at T > 1200 K. Absorption spectra of quenched samples after LH to ~1000 K and 1200 K are identical to the corresponding before-heating spectra indicating a reversibility of the temperature-induced transformations. The spectrum collected after heating to 1625 K is slightly less absorbing probably due to stresses annealing at high temperature or a slight sample defocusing due to the optics warming over the heating cycle.

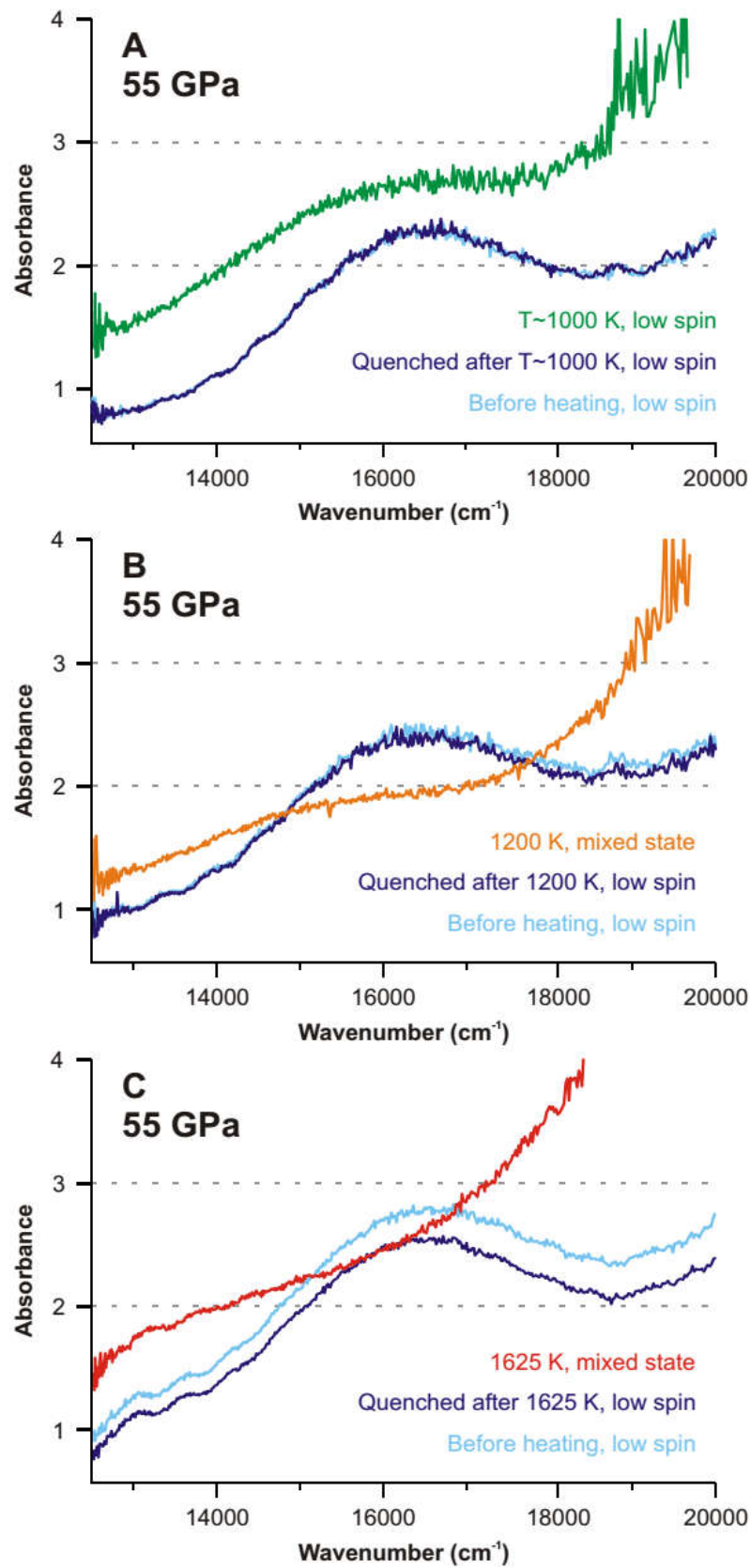

**Fig. 4.** High temperature siderite absorption spectra at 55 GPa. Light and dark blue curves are spectra measured before heating and after heating to corresponding temperature. Green curve in **A**, orange in **B**, and red in **C** are high temperature absorption data at ~1000, 1200, and 1625 K, respectively. Grey dashed lines are guides to the eye. Temperature-induced red-shift of the AE (absorption edge) is evident in **A**, **B**, and **C**. Pressure medium is KCl.

### 73 GPa

Figure 5 shows absorption spectra of siderite at 73 GPa at which pressure the $^1A_{1g} \rightarrow {}^1T_{1g}$ CF band is centered at > 17000 cm$^{-1}$ and was not clearly resolved because of the overlap with the AE and the overall large absorbance reaching ~4, which is close to the detection limit of the spectrometer. The spectra are now dominated by the AE which shows an apparent red-shift with increasing temperature (~ 1cm$^{-1}$/K) making the CF band extremely difficult to follow. One peculiar feature is an increased absorbance at 12500-16000 cm$^{-1}$ at T = 1000 K (Fig. 5A). We attribute this to the intensification, broadening, and red-shift of the $^1A_{1g} \rightarrow {}^1T_{1g}$ band, as in the absorption spectrum of LS siderite at ~1000 K at 55 GPa, where the temperature-enhanced CF band is more apparent (Fig. 4A). Further temperature increase to 1225 K results in the decrease of absorbance at 12500-16000 cm$^{-1}$ (Fig. 5A) indicating, perhaps, an increased population of the HS states which are less absorbing. At 1425 K, however, the red-shifted AE causes a gradual increase in absorbance at 13000-16000 cm$^{-1}$ (Fig. 5B).

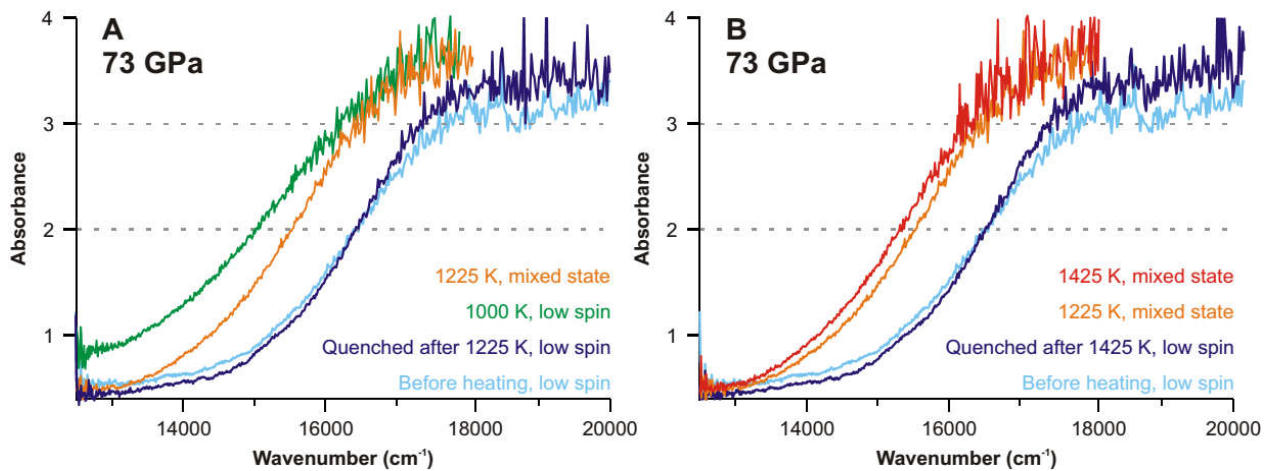

**Fig. 5.** High temperature siderite absorption spectra at 73 GPa. Light and dark blue curves are spectra measured before heating and after heating to corresponding temperature. Green in **A**, orange in **A** and **B**, and red in **B** are absorption spectra at 1000, 1225, and 1425 K, respectively. Temperature-induced red-shift of the AE (absorption edge) is ~ 1 (cm$^{-1}$/K). Grey dashed lines are guides to the eye. Pressure medium is KCl.

### Radiative thermal conductivity of the spin transition zone

Traditionally, the key assumption for quantifying radiative thermal conductivity of a material is that the absorption coefficient is temperature-independent in a wide spectral range (*e.g.* 2000-25000 cm$^{-1}$)[4, 5, 7-9, 11-18, 23, 24, 46]. This allows using room temperature absorption coefficients in the absence of high temperature absorption data. Our results are at odds with this assumption suggesting that a room temperature absorption coefficient may not yield accurate dependence of $k_{rad}$ on temperature. In order to estimate the difference between the classically determined $k_{rad}$ (300 K absorption data) and $k_{rad}$ quantified using high temperature absorption data we extend the 73 GPa spectra (Fig. 5) to < 12500 cm$^{-1}$ using the previously reported near- and mid-IR siderite absorption data[26] and assuming the absorption spectrum at < 12500 cm$^{-1}$ is temperature-independent as it is featureless. In quantifying the absorption coefficients at 300, 1000, 1225, and 1425 K we suppose the sample thickness of 10 μm as we are interested only in relative differences in $k_{rad}$. We may now use the 300, 1000, 1225, and 1425 K absorption coefficients within the traditional approach[8, 9, 24] to understand the magnitude of difference in $k_{rad}$ introduced by assuming temperature-independent absorbance (Fig. 6). Importantly, the curves shown in Figure 6 reflect only relative differences in $k_{rad}$ as a function of spin state revealing the previously ignored ambiguity introduced by the use of a room temperature absorption coefficient. The absolute values are arbitrary because of the uncertainty introduced by the increased overall absorption due to the use of rigid pressure mediums required to isolate the sample from diamonds.

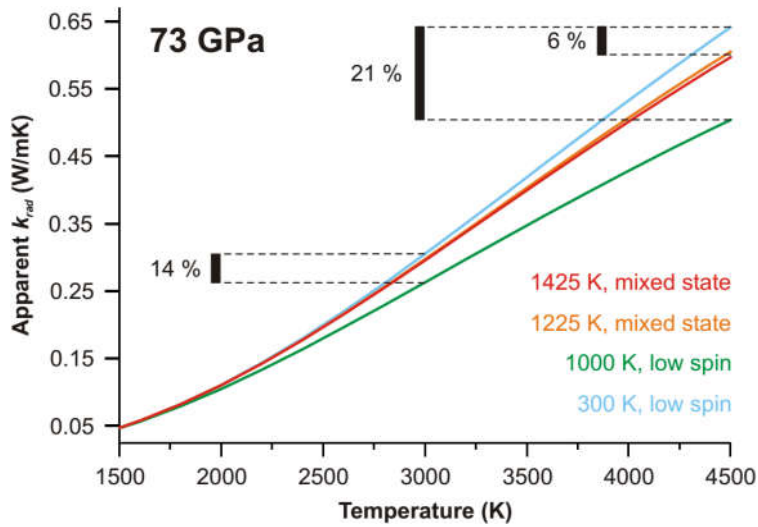

**Fig. 6.** Apparent radiative thermal conductivity ($k_{rad}$) of siderite at 73 GPa calculated using absorption spectra shown in Fig. 5 appended to near-, mid-IR data reported by Lobanov et al. [26] in the 2500-12500 cm$^{-1}$ spectra range (71.5 GPa, 300 K) and assuming absorbance at < 12500 cm$^{-1}$ is temperature-independent. The color refers the curves to the absorption spectra shown in Fig. 5. Maximum differences in $k_{rad}$ are on the order of 21 % at 4500 K.

Moderate differences in $k_{rad}$ calculated using 300 K vs high temperature absorption data peak at the highest temperatures (up to 21 % for the LS state) because of the increased overlap of the siderite absorption features with the blackbody radiation spectrum. Radiative thermal conductivity of a FeO$_6$-bearing mineral with a mixed iron spin state (1225 and 1425 K reference absorption spectra) appears < 6% lower than the curve produced by the 300 K absorption coefficient. This is a direct consequence of siderite low optical absorbance at < 13500-14000 cm$^{-1}$ (Fig. 5). In FP the correction may be larger as its CF bands are located at 5000-15000 cm$^{-1}$ and, therefore, closer to the peak position of the thermal radiation spectrum (Refs.[14, 15, 17]). The previously inferred (from its 300 K absorption coefficient) $k_{rad}$ of FP is very small (< 0.1 W/mK) at the core-mantle boundary (CMB)[11, 12]. Theoretical computations indicate that the fraction of HS states is small (< 0.5) at the base of the lower mantle[47-50]. If such, the temperature-induced red-shift and intensification of the $^1A_{1g} \rightarrow {}^1T_{1g}$ band observed in this study may play an important role in the ability of FP to conduct heat by radiation. Indeed, the green curve in Figure 6 suggests that FP radiative thermal conductivity at CMB temperatures is at least 21 % lower than the previously reported value[11, 12]. Moreover, if the observed constant temperature-induced red-shift of the AE (~ 1cm$^{-1}$/K) (Fig. 5) is an intrinsic feature of a FeO$_6$-octahedra, then the AE in FP may exhibit a similar temperature trend, effectively blocking the visible and near-IR range at CMB temperatures. The latter is consistent with the finite-temperature first-principles molecular dynamic simulations indicating a semimetallic behavior of FP at the base of the mantle[48]. In this scenario, radiative thermal conductivity of FP is negligible at CMB conditions.

**Conclusions**

To summarize, we examined temperature-induced variations in optical properties of low and high spin FeO$_6$ octahedral unit in siderite at 45-73 GPa in the 12500-22500 cm$^{-1}$ spectral range. We find that the absorbance of low spin FeO$_6$ increases with temperature, but the temperature-induced low-to-high spin transition results in an abrupt reduction of optical absorbance of the crystal field bands. The absorption edge, however, shows a continuous red-shift with increasing temperature suggesting its major role in diminishing the ability of FeO$_6$-bearing minerals, such as ferropericlase, to conduct heat by radiation at mantle temperatures. This supports a reduced radiative conductivity scenario in the lower mantle. Additionally, we show that high temperature absorption coefficients are needed to carefully address the mantle's radiative thermal conductivity.

**Acknowledgement**

We are grateful to Tohoku University of Natural History and Konstantin Litasov for providing siderite samples. This work was supported by the NSF Major Research Instrumentation program, NSF EAR-1015239, NSF EAR-1520648 and NSF EAR/IF-1128867, the Army Research Office (56122-CH-H), the Carnegie Institution of Washington and Deep Carbon Observatory. S.S.L. was partly supported by the Ministry of Education and Science of Russian Federation (No. 14B.25.31.0032). Portions of this work were performed at GeoSoilEnviroCARS (Sector 13), Advanced Photon Source (APS), Argonne National Laboratory. GeoSoilEnviroCARS is supported by the National Science Foundation - Earth Sciences (EAR-1128799) and Department of Energy- GeoSciences (DE-FG02-94ER14466). This research used resources of the Advanced Photon Source, a U.S. Department of Energy (DOE) Office of Science User Facility operated for the DOE Office of Science by Argonne National Laboratory under Contract No. DE-AC02-06CH11357. Zachary M. Geballe is thanked for his comments on the earlier version of this manuscript.

# References

1. Dubuffet, F., Yuen, D. A., Rainey, E. S. G. Controlling thermal chaos in the mantle by positive feedback from radiative thermal conductivity. *Nonlinear Process. Geophys.,* **9,** 311-323 (2002).

2. Hofmeister, A. M. Mantle values of thermal conductivity and the geotherm from phonon lifetimes. *Science,* **283,** 1699-1706 (1999).

3. Stacey, F. D. & Davis, P. M. *Physics of the Earth*, (Cambridge University Press, Cambridge ; New York, 2008).

4. Clark, S. P. Radiative transfer in the Earth's mantle. *Eos (formerly Trans. Am. Geophys. Union),* **38,** 931-938 (1957).

5. Hofmeister, A. M. Enhancement of radiative transfer in the upper mantle by $OH^-$ in minerals. *Phys. Earth Planet. Inter.,* **146,** 483-495 (2004).

6. Gibert, B., Schilling, F. R., Gratz, K., Tommasi, A. Thermal diffusivity of olivine single crystals and a dunite at high temperature: Evidence for heat transfer by radiation in the upper mantle. *Phys. Earth Planet. Inter.,* **151,** 129-141 (2005).

7. Clark, S. P. Absorption spectra of some silicates in the visible and near Infrared. *Am. Mineral.,* **42,** 732-742 (1957).

8. Shankland, T. J., Nitsan, U., Duba, A. G. Optical absorption and radiative heat transport in olivine at high temperature. *Journal of Geophysical Research,* **84,** 1603-1610 (1979).

9. Fukao, Y., Mizutani, H., Uyeda, S. Optical absorption spectra at high temperatures and radiative thermal conductivity of olivines. *Phys. Earth Planet. Inter.,* **1,** 57-62 (1968).

10. Hofmeister, A. M., Branlund, J. M., Pertermann, M. in *Treatise on Geophysics*, 543-577, (Amsterdam, 2007).

11. Goncharov, A. F., Haugen, B. D., Struzhkin, V. V., Beck, P., Jacobsen, S. D. Radiative conductivity in the Earth's lower mantle. *Nature,* **456,** 231-234 (2008).

12. Goncharov, A. F., Beck, P., Struzhkin, V. V., Haugen, B. D., Jacobsen, S. D. Thermal conductivity of lower-mantle minerals. *Phys. Earth Planet. Inter.,* **174,** 24-32 (2009).

13. Keppler, H. & Smyth, J. R. Optical and near infrared spectra of ringwoodite to 21.5 GPa: Implications for radiative heat transport in the mantle. *Am. Mineral.,* **90,** 1209-1212 (2005).

14. Keppler, H., Kantor, I., Dubrovinsky, L. S. Optical absorption spectra of ferropericlase to 84 GPa. *Am. Mineral.,* **92,** 433-436 (2007).

15. Goncharov, A. F., Struzhkin, V. V., Jacobsen, S. D. Reduced radiative conductivity of low-spin (Mg,Fe)O in the lower mantle. *Science,* **312,** 1205-1208 (2006).

16. Keppler, H., Dubrovinsky, L. S., Narygina, O., Kantor, I. Optical absorption and radiative thermal conductivity of silicate perovskite to 125 Gigapascals. *Science,* **322,** 1529-1532 (2008).

17. Goncharov, A. F*., et al.* Effect of composition, structure, and spin state on the thermal conductivity of the Earth's lower mantle. *Phys. Earth Planet. Inter.,* **180,** 148-153 (2010).

18. Goncharov, A. F*., et al.* Experimental study of thermal conductivity at high pressures: Implications for the deep Earth's interior. *Phys. Earth Planet. Inter.,* **247,** 11-16 (2015).

19. Mao, H. K. & Bell, P. M. Electrical conductiviy and the red shift of absorption in olivine and spinel at high pressure. *Science,* **176,** 403-405 (1972).

20. Taran, M. N*., et al.* Spectroscopic studies of synthetic and natural ringwoodite, γ-(Mg, Fe)$_2$SiO. *Phys. Chem. Miner.,* **36,** 217-232 (2009).

21. Sung, C. M., Singer, R. B., Parkin, K. M., Burns, R. G. Temperature dependence of $Fe^{2+}$ crystal field spectra: Implications to mineralogical mapping of planetary surfaces. VIII Lunar and Planetary Science Conference; 1977; 1977. p. 1063-1079.

22. Ullrich, K., Langer, K., Becker, K. D. Temperature dependence of the polarized electronic absorption spectra of olivines. Part I - fayalite. *Phys. Chem. Miner.,* **29,** 409-419 (2002).

23. Thomas, S. M., Bina, C. R., Jacobsen, S. D., Goncharov, A. F. Radiative heat transfer in a hydrous mantle transition zone. *Earth Planet. Sci. Lett.,* **357,** 130-136 (2012).

24. Murakami, M*., et al.* High-pressure radiative conductivity of dense silicate glasses with potential implications for dark magmas. *Nature Comm.,* **5,** 5428 (2014).

25. Burns, R. G. *Mineralogical applications of crystal field theory*, (Cambridge University Press, U.K., 1993).

26. Lobanov, S. S., Goncharov, A. F., Litasov, K. D. Optical properties of siderite ($FeCO_3$) across the spin transition: Crossover to iron-rich carbonates in the lower mantle. *Am. Mineral.,* **100,** 1059-1064 (2015).

27. Farfan, G., Wang, S. B., Ma, H. W., Caracas, R., Mao, W. L. Bonding and structural changes in siderite at high pressure. *Am. Mineral.,* **97,** 1421-1426 (2012).

28. Lavina, B., et al. Siderite at lower mantle conditions and the effects of the pressure-induced spin-pairing transition. *Geophys. Res. Lett.,* **36,** L23306 (2009).

29. Mattila, A., et al. Pressure induced magnetic transition in siderite FeCO3 studied by x-ray emission spectroscopy. *J. Phys. Cond. Matter,* **19,** (2007).

30. Cerantola, V., et al. High-pressure spectroscopic study of siderite (FeCO3) with a focus on spin crossover. *Am. Mineral.,* **100,** 2670-2681 (2015).

31. Lin, J. F., Speziale, S., Mao, Z., Marquardt, H. Effects of the electronic spin transitions of iron in lower mantle minerals: iImplications for deep mantle geophysics and geochemistry. *Rev. Geophys.,* **51,** 244-275 (2013).

32. Marquardt, H., Speziale, S., Reichmann, H. J., Frost, D. J., Schilling, F. R. Single-crystal elasticity of (Mg0.9Fe0.1)O to 81 GPa. *Earth Planet. Sci. Lett.,* **287,** 345-352 (2009).

33. Crowhurst, J. C., Brown, J. M., Goncharov, A. F., Jacobsen, S. D. Elasticity of (Mg,Fe)O through the spin transition of iron in the lower mantle. *Science,* **319,** 451-453 (2008).

34. Speziale, S., et al. Iron spin transition in Earth's mantle. *Proc. Natl. Acad. Sci. U.S.A.,* **102,** 17918-17922 (2005).

35. Mao, H. K., Bell, P. M., Shaner, J. W., Steinberg, D. J. Specific volume measurements of Cu, Mo, Pd, and Ag and calibration of the ruby $R_1$ fluorescence pressure gauge from 0.06 to 1 Mbar. *J. Appl. Phys.,* **49,** 3276-3283 (1978).

36. Akahama, Y. & Kawamura, H. Pressure calibration of diamond anvil Raman gauge to 310 GPa. *J. Appl. Phys.,* **100,** (2006).

37. McWilliams, R. S., Dalton, D. A., Konopkova, Z., Mahmood, M. F., Goncharov, A. F. Opacity and conductivity measurements in noble gases at conditions of planetary and stellar interiors. *Proc. Natl. Acad. Sci. U.S.A.,* **112,** 7925-7930 (2015).

38. Goncharov, A. F., et al. Laser heating in diamond anvil cells: developments in pulsed and continuous techniques. *J. Synchrotron Radiat.,* **16,** 769-772 (2009).

39. Benedetti, L. R., Antonangeli, D., Farber, D. L., Mezouar, M. An integrated method to determine melting temperatures in high-pressure laser-heating experiments. *Appl. Phys. Lett.,* **92,** (2008).

40. Taran, M. N. & Langer, K. Electronic absorption spectra of $Fe^{2+}$ ions in oxygen-based rock-forming minerals at temperatures between 297 and 600 K. *Phys. Chem. Miner.,* **28,** 199-210 (2001).

41. Liu, J., Lin, J. F., Mao, Z., Prakapenka, V. B. Thermal equation of state and spin transition of magnesiosiderite at high pressure and temperature. *Am. Mineral.,* **99,** 84-93 (2014).

42. Boulard, E.*, et al.* New host for carbon in the deep Earth. *Proc. Natl. Acad. Sci. U.S.A.,* **108,** 5184-5187 (2011).

43. Boulard, E.*, et al.* Experimental investigation of the stability of Fe-rich carbonates in the lower mantle. *Journal of Geophysical Research,* **117,** (2012).

44. Taran, M. N., Langer, K., Platonov, A. N., Indutny, V. V. Optical absorption investigation of $Cr^{3+}$ ion-bearing minerals in the temperature range 77-797 K. *Phys. Chem. Miner.,* **21,** 360-372 (1994).

45. Ullrich, K., Ott, O., Langer, K., Becker, K. D. Temperature dependence of the polarized electronic absorption spectra of olivines. Part II - Cobalt-containing olivines. *Phys. Chem. Miner.,* **31,** 247-260 (2004).

46. Hofmeister, A. M. Dependence of diffusive radiative transfer on grain-size, temperature, and Fe-content: Implications for mantle processes. *J. Geodyn.,* **40,** 51-72 (2005).

47. Tsuchiya, T., Wentzcovitch, R. M., da Silva, C. R. S., de Gironcoli, S. Spin transition in magnesiowustite in Earth's lower mantle. *Phys. Rev. Lett.,* **96,** 198501 (2006).

48. Holmstrom, E. & Stixrude, L. Spin crossover in ferropericlase from first-principles molecular dynamics. *Phys. Rev. Lett.,* **114,** 117202 (2015).

49. Lyubutin, I. S.*, et al.* Quantum critical point and spin fluctuations in lower-mantle ferropericlase. *Proc. Natl. Acad. Sci. U.S.A.,* **110,** 7142-7147 (2013).

50. Sturhahn, W., Jackson, J. M., Lin, J. F. The spin state of iron in minerals of Earth's lower mantle. *Geophys. Res. Lett.,* **32,** L12307 (2005).